\documentclass[preprint,12pt]{elsarticle}

\usepackage{amsmath}
\usepackage{amssymb}
\usepackage{bm}
\usepackage{graphicx}
\usepackage{subcaption}
\usepackage{multirow}
\usepackage{booktabs}
\usepackage{siunitx}
\usepackage{xcolor}
\usepackage{hyperref}

\journal{Materials Today Physics}

\begin{document}

\begin{frontmatter}

\title{Emergent Noncoplanar Spin Chirality through Magnetic and Nonmagnetic Substitution in the Kagome Antiferromagnet Mn$_3$Sn}

\author[snb]{Anupam Barik}
\author[snb]{Achintya Low\fnref{fn1}}
\author[snb]{Susanta Ghosh\fnref{fn2}}
\author[iit]{Kapildeb Dolui}
\author[snb]{Setti Thirupathaiah\corref{cor1}}

\cortext[cor1]{Corresponding author}
\ead{setti@bose.res.in}

\address[snb]{Department of Condensed Matter and Materials Physics,
S.~N. Bose National Centre for Basic Sciences,
Kolkata-700106, India}

\address[iit]{Department of Physics,
Indian Institute of Technology Tirupati,
Tirupati-517619, India}

\fntext[fn1]{Present address: Department of Physics,
Ben-Gurion University of the Negev,
Be'er-Sheva 84105, Israel.}

\fntext[fn2]{Present address: Department of Chemistry,
Indian Institute of Science Education and Research Bhopal,
Bhopal 462066, India.}

\begin{abstract}
Controlling noncoplanar spin textures in centrosymmetric magnets remains
a central challenge for realizing emergent Berry-phase phenomena. Here,
we demonstrate that magnetic (Cr) and nonmagnetic (Cu) substitution at
the Mn site provide distinct routes to engineer spin chirality and
topological transport in the kagome antiferromagnet Mn$_3$Sn. Cr
substitution enhances low-temperature spin canting, stabilizing a dense
noncoplanar texture that yields a pronounced topological Hall effect
(THE) at low temperatures. In contrast, Cu substitution drives a
spin-reorientation transition accompanied by an easy-axis rotation,
creating a broad temperature window (150--208 K) where competing magnetic
anisotropies stabilize noncoplanar spin textures and generate a large
THE. A concurrent carrier-type crossover signals an underlying
electronic-structure reconstruction. These results reveal that exchange
enhancement and exchange dilution represent two fundamentally different
pathways for tuning real-space Berry curvature, establishing chemical
substitution as a powerful strategy for engineering topological transport
in kagome antiferromagnets.
\end{abstract}

\begin{highlights}
\item Magnetic and nonmagnetic substitution provide distinct routes to
control spin chirality in Mn$_3$Sn.
\item Cr substitution enhances low-temperature noncoplanar spin textures
and the topological Hall effect.
\item Cu substitution induces a spin-reorientation transition and a broad
intermediate-temperature topological Hall regime.
\item The chiral spin (skyrmion) density and characteristic length scale reveal distinct
noncoplanar textures, with $\lambda_{\mathrm{sk}}\sim3.7$~nm for Cr and
$\sim39$~nm for Cu substitution.
\item Chemical substitution provides a route to engineer real-space Berry
curvature and tunable chiral spin textures in kagome antiferromagnets.
\end{highlights}


\end{frontmatter}

\section{Introduction}

The interplay between nontrivial band topology and magnetism has emerged as a central theme in condensed matter physics, enabling the realization of exotic quantum states and unconventional transport phenomena~\cite{Jiang2021unconventional, Wang2021charge,Moore2022topological, Fu2021quantum,Dutta2025electronic, Roychowdhury2023interplay}. In magnetic systems, the breaking of time-reversal symmetry generates momentum-space Berry curvature~\cite{Sinitsyn2005disorder,xiao2010berry,nagaosa2010anomalous}, giving rise to an intrinsic anomalous Hall effect (AHE)~\cite{kakati2024berry,thakur2020intrinsic,zeng2022large,liu2018giant}, even in materials with nearly vanishing net magnetization~\cite{hou2023large,Jiang2021unconventional}. Beyond momentum-space topology, noncoplanar spin textures with finite scalar spin chirality, $\chi_{ijk}=\mathbf{S}_i\cdot(\mathbf{S}_j\times\mathbf{S}_k)$, generate a real-space Berry phase that acts as an emergent magnetic field on itinerant electrons, producing an additional topological Hall effect (THE)~\cite{bruno2004topological,onoda2004anomalous,Zheng2021giant}. The coexistence and independent tunability of momentum- and real-space Berry phases make frustrated magnetic materials an ideal platform for exploring emergent topological transport phenomena.

Among such systems, the kagome metals have attracted particular attention because geometrical frustration, Dirac fermions, flat bands, and van Hove singularities naturally promote unconventional magnetic order and topological electronic states~\cite{balents2010spin,kang2020dirac,liu2020orbital,lin2018flatbands,kang2022twofold,kimbell2022challenges}. These characteristics provide an ideal platform for realizing and controlling Berry-phase-driven transport. Among kagome magnets, Mn$_3$Sn has emerged as a prototypical chiral antiferromagnetic Weyl semimetal~\cite{jiao2019signatures,xu2015discovery}. It crystallizes in the hexagonal $P6_3/mmc$ structure, where Mn atoms form stacked kagome layers along the $c$ axis~\cite{yang2017topological,brown1990determination,tomiyoshi1982magnetic}. Above the spin-reorientation transition, the Mn moments adopt an inverse triangular noncollinear antiferromagnetic configuration within the kagome plane~\cite{brown1990determination,nakatsuji2015large,nagamiya1982triangular}. Although the net magnetization is nearly zero, the inverse triangular spin structure breaks time-reversal symmetry and generates a large Berry curvature near the Fermi level~\cite{Li2023field,rout2019field,ghimire2018large}, resulting in an intrinsic AHE comparable to that of ferromagnets~\cite{nagaosa2010anomalous}. Upon cooling below the spin-reorientation temperature ($T_{SR}\sim260 - 265$ K), the system transforms into a spin-spiral state in which the momentum-space Berry curvature is largely suppressed~\cite{liu2017anomalous,nakatsuji2015large,park2018magnetic,low2022tuning}. This strong coupling between magnetic order and Berry curvature makes Mn$_3$Sn an ideal platform for engineering topological transport through controlled modification of its spin texture.

Despite its globally centrosymmetric crystal structure, the absence of inversion symmetry at individual Mn--Mn bond centers permits finite local Dzyaloshinskii--Moriya interactions (DMI)~\cite{low2025topological,ghosh2023tuning,zhang2023out, Liu2023, Liu2023a}, providing a microscopic mechanism for stabilizing noncoplanar spin textures. In this regard, chemical substitution offers an effective route to tune exchange interactions, magnetic anisotropy, carrier concentration, and spin canting. Previous investigations have primarily focused on magnetic dopants such as Fe and Cr, which modify the inverse triangular antiferromagnetic order and enhance weak ferromagnetic components at low temperatures~\cite{Yano2024tuning,Singh2024higher,Hanggai2025complex,Tagami2024magnetic,low2022tuning}. In contrast, the influence of non-magnetic substitution on the stabilization of noncoplanar spin textures and the resulting topological transport remains largely unexplored. A systematic comparison between magnetic and non-magnetic substitution in Mn$_3$Sn is therefore essential for understanding the microscopic mechanisms governing topological transport.

Here, we present a comparative investigation of the magnetic structure and Hall transport in Mn$_3$Sn upon Cr and Cu substitution at the Mn site. We demonstrate that magnetic (Cr) and non-magnetic (Cu) dopants drive fundamentally different magnetic ground states and Hall responses. Cr substitution strengthens exchange interactions and stabilizes an out-of-plane canted spin state at low temperatures, leading to a pronounced topological Hall effect consistent with short-period noncoplanar spin textures. In contrast, Cu substitution weakens the Mn--Mn exchange network and induces a spin-reorientation transition accompanied by an easy-axis rotation, where competing magnetic anisotropies stabilize noncoplanar spin configurations over a broad intermediate-temperature regime, resulting in an enhanced Hall response. Density functional theory calculations reveal distinct magnetic-anisotropy landscapes for the two dopants, demonstrating that the interplay among exchange interactions, magnetic anisotropy, and electronic filling governs the emergence of real-space Berry curvature. Our findings establish a microscopic picture of dopant-controlled topological transport in kagome antiferromagnets and provide a general strategy for engineering Berry-phase phenomena through selective chemical substitution.

\begin{figure}[htbp]
\centering
\includegraphics[width=0.8\linewidth]{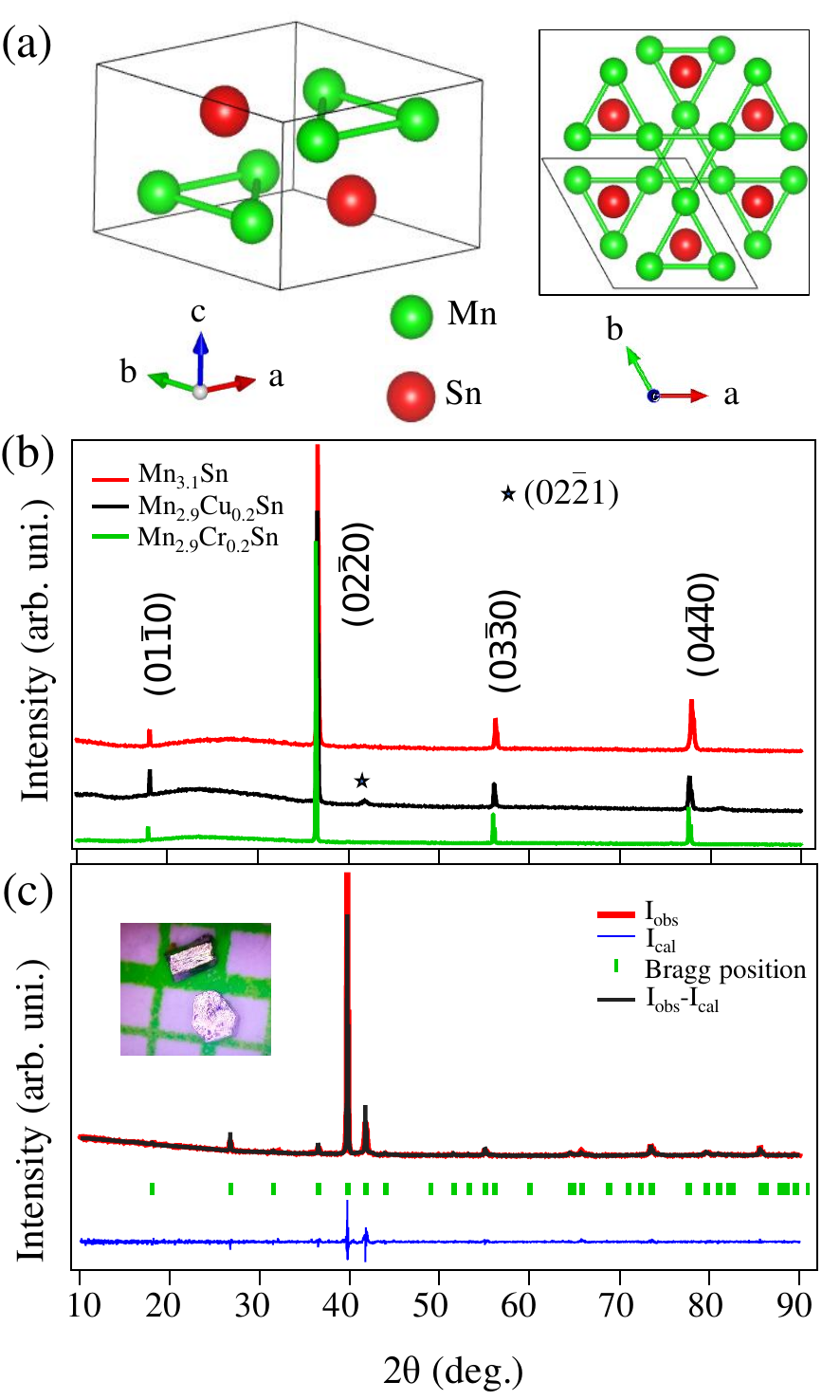}
\caption{(a)~Three-dimensional crystal structure of Mn$_3$Sn (left panel) and its projection onto the $ab$ plane (right panel), highlighting the kagome lattice formed by Mn atoms (green spheres). (b)~X-ray diffraction (XRD) patterns of Mn$_{3.1}$Sn,
Mn$_{2.9}$Cu$_{0.2}$Sn, and Mn$_{2.9}$Cr$_{0.2}$Sn single crystals measured along the $[0l\bar{l}0]$ crystallographic direction. (c)~Powder XRD pattern of crushed Mn$_{3.1}$Sn single crystals overlaid with the Rietveld refinement profile. Owing to the small crystal size, a weak-intensity peak corresponding to the $[02\bar{2}1]$ Bragg reflection is observed for Mn$_{2.9}$Cu$_{0.2}$Sn.}
\label{fig1}
\end{figure}

\begin{figure*}[htbp]
\centering
\includegraphics[width=\linewidth]{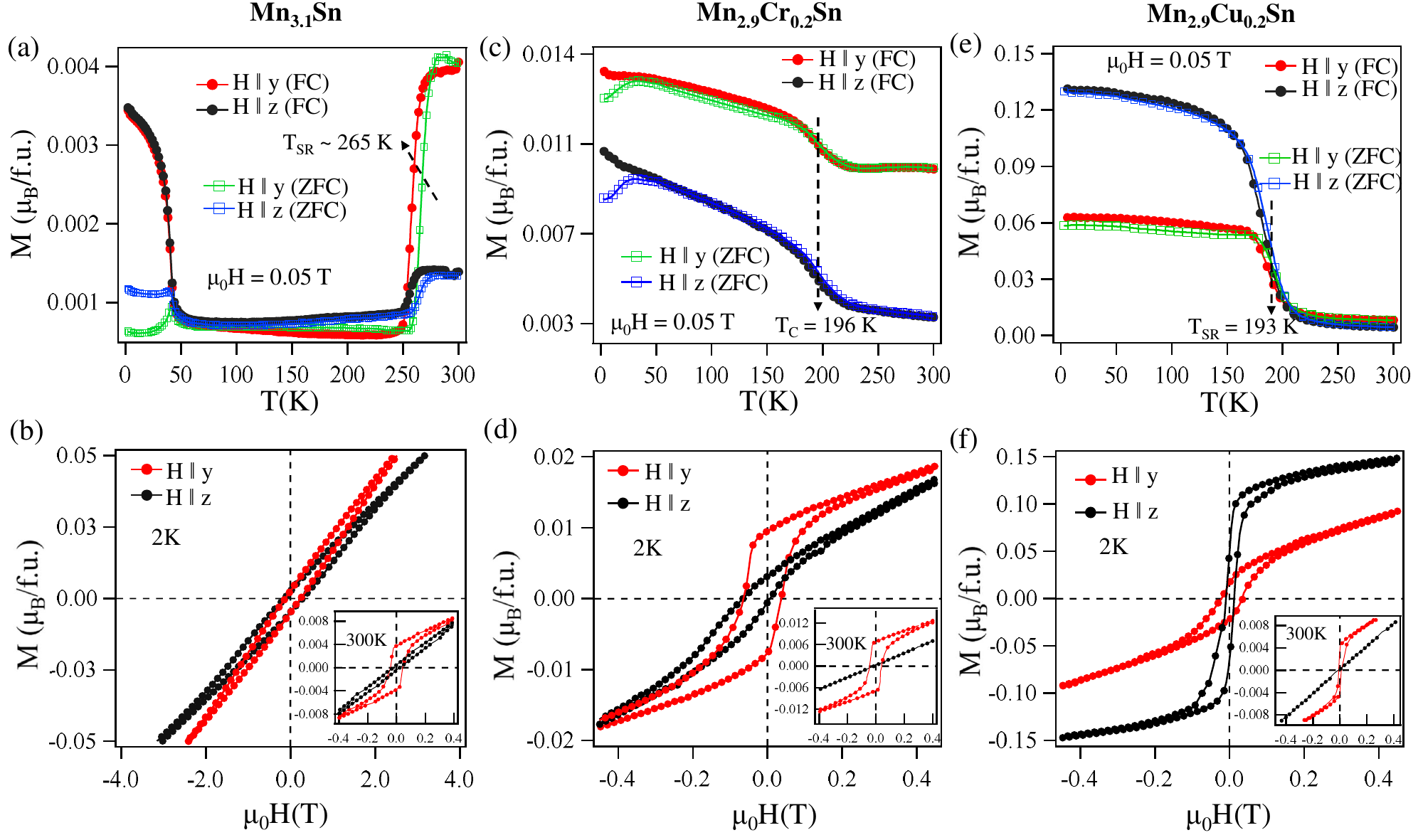}
\caption{Temperature-dependent magnetization [$M(T)$] measured in the range 2--300\,K for (a)~Mn$_{3.1}$Sn, (c)~Mn$_{2.9}$Cr$_{0.2}$Sn, and
(e)~Mn$_{2.9}$Cu$_{0.2}$Sn under an applied field of 0.05\,T in both zero-field-cooled (ZFC) and field-cooled (FC) modes, with the field applied both parallel ($H\parallel z$) and perpendicular ($H\parallel y $) to the crystallographic $c$-axis. Field-dependent isothermal magnetization [$M(H)$] measured at selected temperatures for (b)~Mn$_{3.1}$Sn, (d)~Mn$_{2.9}$Cr$_{0.2}$Sn, and (f)~Mn$_{2.9}$Cu$_{0.2}$Sn measured at 2\,K.  Corresponding data measured at 300 K is shown in the respective insets.}
\label{fig2}
\end{figure*}

\section{Methods}
\subsection{Experimental Details}

Single crystals of Mn$_3$Sn and its substituted variants were grown by the self-flux method~\cite{canfield1992growth,sung2018magnetic,zhang2023out}.
High-purity elemental powders of Mn (Alfa Aesar, 99.995\%), Cu or Cr
(Alfa Aesar, 99.99\%), and Sn (Alfa Aesar, 99.998\%) were weighed in the molar ratio $(7-x):x:3$,
where $x$ denotes the nominal substitution level. The mixture was loaded into an alumina crucible,
sealed in an evacuated quartz ampoule, and heated to $1000\,^\circ\mathrm{C}$ to ensure complete melting and homogenization.
The melt was subsequently cooled to $900\,^\circ\mathrm{C}$ at a rate of $2^{\circ}C/h$ to promote single-crystal nucleation.
At $900\,^\circ\mathrm{C}$, the ampoule was removed from the furnace and immediately centrifuged to separate the excess Sn flux,
yielding shiny hexagonal rod-shaped crystals with typical dimensions of approximately $1\times0.5\times0.5\,\mathrm{mm}^3$.

Phase purity and structural characterization were carried out by x-ray diffraction (XRD) on a Rigaku SmartLab 9\,kW diffractometer with Cu $K_\alpha$ radiation. Elemental compositions were determined by energy-dispersive x-ray spectroscopy (EDXS). The EDXS measurements (not shown) confirm the elemental compositions of Mn$_{3.1}$Sn, Mn$_{2.9}$Cu$_{0.2}$Sn, and Mn$_{2.9}$Cr$_{0.2}$Sn to within an experimental uncertainty of approximately 2\%. The slight excess of Mn relative to the ideal stoichiometry is commonly observed in flux-grown Mn$_3$Sn crystals and is attributed to the stabilization of interstitial Mn during growth~\cite{sung2018magnetic}. The doping concentration of 0.2 was chosen in this study based on our previous systematic study of Fe-doped Mn$_3$Sn~\cite{low2025topological}, which revealed a maximum topological Hall response at this composition. Accordingly, the same doping level was employed here to systematically examine the influence of different dopant elements on the magnetic and magnetotransport properties. Electrical transport, Hall effect, and magnetization measurements were performed using a 9\,T Physical Property Measurement System (PPMS, Quantum Design). To eliminate artifacts from longitudinal resistivity arising from possible voltage-probe misalignment, the Hall resistivity was extracted from the antisymmetrized component, $\rho_{zx} = \frac{\rho_{zx}(+\mu_0 H) - \rho_{zx}(-\mu_0 H)}{2}$.

\begin{figure*}[htbp]
\centering
\includegraphics[width=1\linewidth]{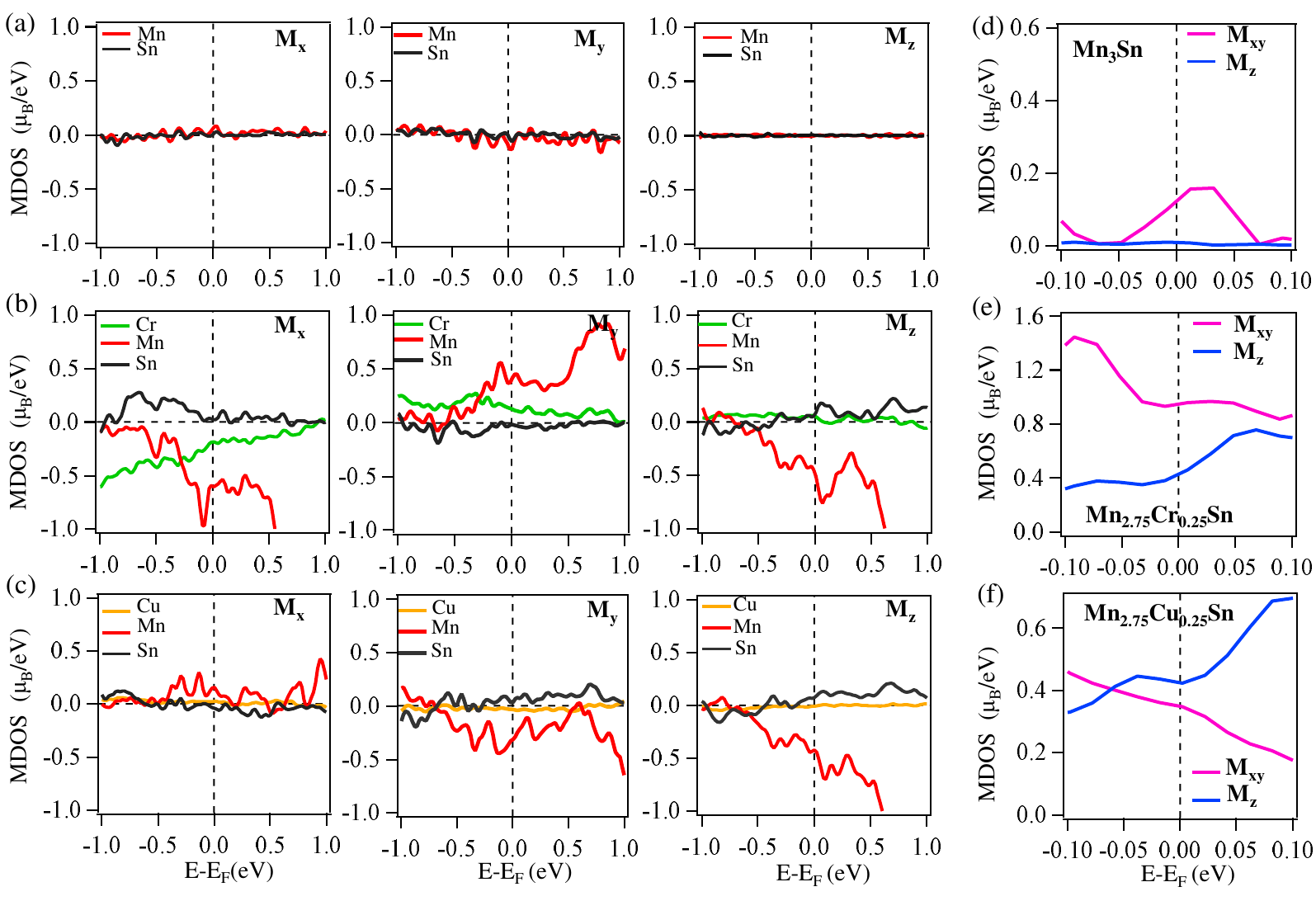}
\caption{Magnetization density of states (MDOS) calculated from DFT for pristine and doped Mn$_3$Sn systems. The MDOS, projected onto the Mn, Sn, Cr and Cu atoms, are resolved for (a)~$\text{Mn}_3\text{Sn}$, (b)~$\text{Mn}_{2.75}\text{Cu}_{0.25}\text{Sn}$, and     (c)~$\text{Mn}_{2.75}\text{Cr}_{0.25}\text{Sn}$ along the Cartesian $x$ ($M_x$), $y$ ($M_y$), and $z$ ($M_z$) axes using the Pauli spin matrices. Panels (d)–(f) compare the net in-plane MDOS magnitude, $M_{xy} = \sqrt{M_x^2 + M_y^2}$, with the out-of-plane component ($M_z$) as a function of energy for $\text{Mn}_3\text{Sn}$, $\text{Mn}_{2.75}\text{Cr}_{0.25}\text{Sn}$, and $\text{Mn}_{2.75}\text{Cu}_{0.25}\text{Sn}$, respectively.}
\label{fig3}
\end{figure*}

\subsection{First-Principles Calculations}

Electronic structure calculations were performed within the framework of density functional theory (DFT) using the generalized gradient approximation (GGA) in the Perdew--Burke--Ernzerhof (PBE)~\cite{Perdew1996} form for the exchange-correlation functional, as implemented in the Vienna \textit{Ab initio} Simulation Package (VASP)~\cite{Kresse1996}. The projector augmented-wave (PAW) method~\cite{Kresse1999} was employed to describe the core electrons. The electronic wave functions were expanded in a plane-wave basis set with a kinetic-energy cutoff of 600~eV. The experimentally determined lattice constants were used for both pristine and doped Mn$_3$Sn structures, while all atomic positions were fully relaxed until the residual forces on each atom were less than 0.04~eV/\AA. The resulting relaxed structures were subsequently used to calculate the electronic band structures and density of states (DOS), with spin-orbit coupling (SOC) included. To account for the on-site Coulomb interactions among the localized $3d$ electrons, the Hubbard $U$ correction was employed within the Liechtenstein formalism~\cite{Liechtenstein1995}. The parameters $U=4.0$~eV and $J=0.5$~eV were used for the Mn $3d$ orbitals, while $U=3.0$~eV was used for both Cr and Cu $3d$ orbitals.

Doping was modeled by substituting two Mn atoms with dopant atoms (Cr or Cu) in a $2\times2\times1$ supercell. This configuration corresponds to a doping concentration of approximately 8\%, which is close to the experimentally determined concentration of about 7\%. The chosen supercell size is sufficient to capture the local structural and electronic effects induced by the dopant atoms while maintaining a reasonable computational cost. A Monkhorst--Pack $k$-point mesh of $4\times4\times10$ was used for the self-consistent (SCF) calculations, while a denser $6\times6\times14$ mesh was employed for the electronic density of states (DOS) calculations.
\section{Results and Discussion}

\subsection{Structural Properties}

The crystal structure of Mn$_3$Sn is schematically shown in Fig.~\ref{fig1}(a). Mn$_3$Sn crystallizes in the hexagonal $\mathrm{Ni_3Sn}$-type structure with space group $P6_3/mmc$, comprising kagome layers of Mn atoms stacked along the $c$-axis and separated by Sn atoms. Fig.~\ref{fig1}(b) presents the x-ray diffraction (XRD) patterns of Mn$_{3.1}$Sn, Mn$_{2.9}$Cu$_{0.2}$Sn, and Mn$_{2.9}$Cr$_{0.2}$Sn single crystals. The diffraction peaks can be exclusively indexed to the $(0l\bar{l}0)$ reflections, confirming the high crystallographic quality and preferred orientation of the as-grown single crystals. The powder XRD pattern obtained from crushed Mn$_{3.1}$Sn single crystals, together with the corresponding Rietveld refinement as shown in Fig.~\ref{fig1}(c), confirms the phase purity and yields refined lattice parameters of $a=b=5.678(4)$~\AA\ and $c=4.536(2)$~\AA, which is in excellent agreement with previously reported values for Mn$_3$Sn~\cite{nakatsuji2015large,tomiyoshi1982magnetic}. No secondary phases or impurity reflections are observed for any of the investigated compositions. The preservation of the hexagonal crystal structure across all compositions, together with the absence of detectable phase segregation, indicates that both Cu and Cr are incorporated at the Mn sites without disrupting the underlying kagome lattice framework.

\subsection{Magnetic Properties}

Temperature-dependent magnetization [$M(T)$] measurements of Mn$_{3.1}$Sn, Mn$_{2.9}$Cr$_{0.2}$Sn, and Mn$_{2.9}$Cu$_{0.2}$Sn are presented in Figs.~\ref{fig2}(a), ~\ref{fig2}(c), and ~\ref{fig2}(e), respectively.  Field-dependent isothermal magnetization [$M(H)$] curves of Mn$_{3.1}$Sn, Mn$_{2.9}$Cr$_{0.2}$Sn, and Mn$_{2.9}$Cu$_{0.2}$Sn are presented in Figs.~\ref{fig2}(b), ~\ref{fig2}(d), and ~\ref{fig2}(f), respectively. The data were taken with the magnetic field applied both parallel ($H\parallel z$) and perpendicular ($H\parallel y$) to the crystallographic $c$-axis under zero-field-cooled (ZFC) and field-cooled (FC) conditions. As shown in Fig.~\ref{fig2}(a), Mn$_{3.1}$Sn exhibits a pronounced decrease in magnetization at the spin-reorientation temperature, $T_\mathrm{SR}\approx265$ K, corresponding to the transition from the high-temperature inverse triangular noncollinear antiferromagnetic (AFM) state to the low-temperature spin-spiral AFM state. A finite magnetization persists above $T_\mathrm{SR}$ due to weak ferromagnetism originating from the slight distortion of the kagome Mn lattice~\cite{low2022tuning,park2025nominal}. Upon further cooling, a spin-glass-like state emerges below $\sim50$ K, reflecting the intrinsic magnetic frustration of the kagome lattice~\cite{sung2018magnetic,park2025nominal}. The field-dependent isothermal magnetization measured at 300 K [inset in Fig.~\ref{fig2}(b)] exhibits a small hysteresis for $H\parallel y$, consistent with the presence of a weak in-plane ferromagnetic component. At 2 K, the isothermal magnetization curve for $H\parallel z$ display spin-glass-like characteristics.

On the other hand, Cr substitution significantly modifies the magnetic response. As shown in Fig.~\ref{fig2}(c), Mn$_{2.9}$Cr$_{0.2}$Sn exhibits a ferrimagnetic-like anomaly near 196 K together with a pronounced bifurcation between the zero-field-cooled (ZFC) and field-cooled (FC) magnetization curves below $\sim30$ K, indicating a suppression of the spin-glass freezing temperature relative to the parent compound~\cite{low2022tuning,park2025nominal,Feng2006glassy}. The isothermal magnetization measured at 2 K [Fig.~\ref{fig2}(d)] reveals ferrimagnetic-like transition with a spontaneous magnetization of $M_\mathrm{sp}\approx0.01~\mu_\mathrm{B}$/f.u for $H\parallel y$. Notably, a strong in-plane magnetic anisotropy is retained throughout the temperature range investigated, with the in-plane magnetization remaining approximately 1.5--2 times larger than its out-of-plane counterpart. These results indicate that Cr substitution predominantly modifies the local magnetic exchange interactions while preserving the in-plane easy axis characteristic of pristine Mn$_{3.1}$Sn. These observations are consistent with previous studies on the similar system~\cite{Yano2024tuning}

\begin{figure*}[htbp]
\centering
\includegraphics[width=0.9\linewidth]{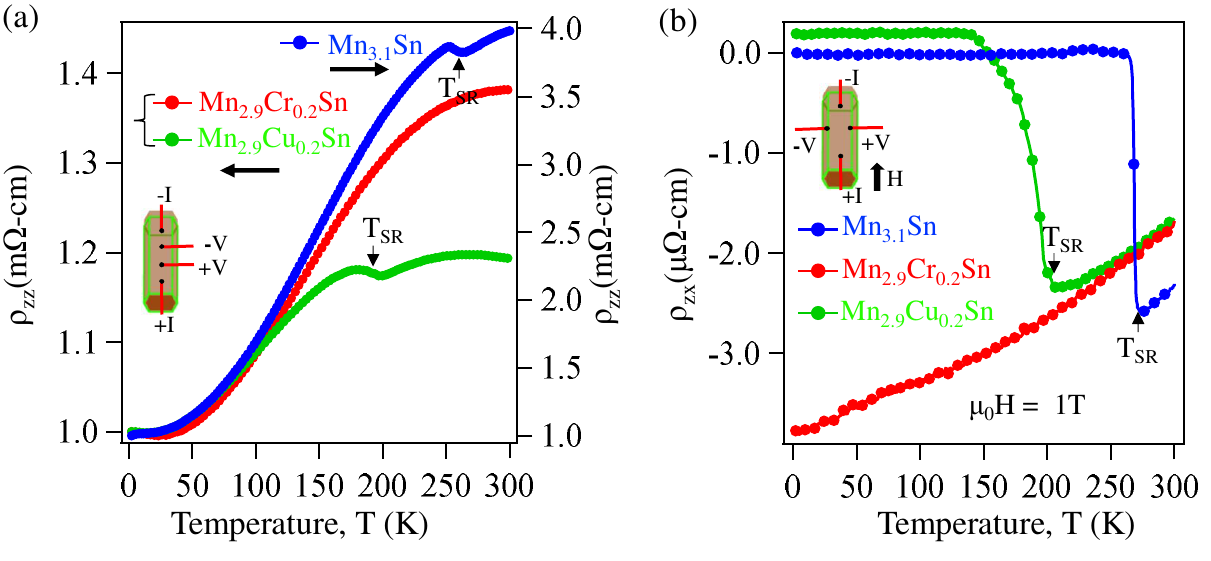}
\caption{(a) Temperature-dependent out-of-plane longitudinal electrical resistivity ($\rho_{zz}$) measured for Mn$_{3.1}$Sn, Mn$_{2.9}$Cr$_{0.2}$Sn, and Mn$_{2.9}$Cu$_{0.2}$Sn over the range 2--300\,K. Arrows mark the spin-reorientation transition temperatures. (b)~Temperature dependence of the Hall resistivity ($\rho_{zx}$) for the same set of samples measured under an applied magnetic field of 1\,T. Inset in (a) and (b) show the measuring geometry of longitudinal resistivity and Hall resistivity, respectively.}
\label{fig4}
\end{figure*}

In contrast, Cu substitution produces a qualitatively different magnetic response to the Cr substitution. As shown in Fig.~\ref{fig2}(e), the magnetization of Mn$_{2.9}$Cu$_{0.2}$Sn is nearly an order of magnitude larger than that of Mn$_{2.9}$Cr$_{0.2}$Sn, demonstrating that nonmagnetic Cu strongly perturbs the inverse triangular antiferromagnetic order and significantly enhances spin canting around the Cu dopant sites. Consequently, the spin-reorientation transition is suppressed to $T_\mathrm{SR}\approx193$ K. More importantly, Cu substitution induces a clear reorientation of the magnetic easy axis across $T_\mathrm{SR}$. That means,  the system evolves from an in-plane easy axis above $T_\mathrm{SR}$ to a predominantly out-of-plane easy axis below $T_\mathrm{SR}$~\cite{gazzah2025doping}, in sharp contrast to the behavior of Mn$_{2.9}$Cr$_{0.2}$Sn. The isothermal $M(H)$ data [Figs.~\ref{fig2}(d) and~\ref{fig2}(f)] further corroborate this picture. At room temperature, the magnetic response of Mn$_{2.9}$Cu$_{0.2}$Sn [see inset of Fig.~\ref{fig2}(f)] resembles that of the parent compound, with the easy axis in the kagome plane. However, at 2 K, a pronounced out-of-plane ferromagnetic component is evident, with a spontaneous magnetization of $\sim 0.1\, \mu_\mathrm{B}$/f.u. for $H\parallel z$ which is significantly enhanced compared to Mn$_{2.9}$Cr$_{0.2}$Sn ($\sim 0.002\,\mu_\mathrm{B}$/f.u). The easy-axis flipping observed in the Cu-doped sample is entirely absent in the Cr-doped system, establishing a fundamental distinction between the two substitution mechanisms.

To better understand the experimental magnetization data, we calculated the magnetization-density of states (MDOS) along the $x$, $y$, and $z$ Cartesian axes for the non-collinear antiferromagnetic (AFM) spin arrangement using the density functional theory.
Here, the MDOS is defined as $ M_i(E) = \text{Tr}\big[\rho(E) \cdot \sigma_i\big]$, where $i \in \{x, y, z\}$
corresponds to the respective Pauli spin matrix ($\sigma_x, \sigma_y, \sigma_z$) and $\rho(E)$ is the density matrix.
This formulation provides crucial microscopic insights into the anisotropic magnetic behavior of $\text{Mn}_3\text{Sn}$ and its doped variants. A co-planar triangular AFM spin orientation is adopted by pristine $\text{Mn}_3\text{Sn}$.

Fig.~\ref{fig3} presents the projected MDOS onto the atomic species Mn, Sn, Cr, and
Cu, plotted as a function of energy and resolved along the Cartesian $x$, $y$, and $z$ magnetization directions.
For pristine $\text{Mn}_3\text{Sn}$ [see Fig.~\ref{fig3}(a)], the MDOS near the Fermi level ($E_F$) is
dominated by the in-plane components ($M_x$ and $M_y$), with a negligible out-of-plane contribution ($M_z$).
The substitution of two Cr atoms for Mn in the $2 \times 2 \times 1$ supercell of pristine $\text{Mn}_3\text{Sn}$, corresponding
to a composition of $\text{Mn}_{2.75}\text{Cr}_{0.25}\text{Sn}$ [see Figs.~\ref{fig3}(b) and  ~\ref{fig3}(e)], leads to a significant
 enhancement of the MDOS near the Fermi level. The in-plane components clearly dominate over the out-of-plane contribution, confirming that the easy-plane anisotropy is preserved within the kagome plane.

In contrast, Cu doped Mn$_3$Sn, $i.e.$ $\text{Mn}_{2.75}\text{Cu}_{0.25}\text{Sn}$ [see Figs.~\ref{fig3}(c) and ~\ref{fig3}(f)] shows a substantial redistribution of MDOS, with the out-of-plane component (M$_z$) exceeding the in-plane contribution near the Fermi level. This directly indicates a reversal of magnetic anisotropy, consistent with the experimentally observed easy-axis rotation. Since Cu is essentially nonmagnetic in this environment, its direct contribution to the net magnetic moment is negligible. Instead, the anisotropy inversion originates from band-filling effects and the resulting redistribution of spin–orbit-coupled Mn $3d$ states near $E_F$.
The directional dependence of the magnetic PDOS further indicates the emergence of a noncollinear spin structure, where different spin channels dominate along orthogonal crystallographic directions. These results demonstrate that chemical substitution modifies the magnetic anisotropy and spin texture through the changes in the electronic structure near the Fermi level, providing a microscopic origin for the observed evolution of magnetic properties in these systems.

\begin{figure*}[htbp]
\centering
\includegraphics[width=\linewidth]{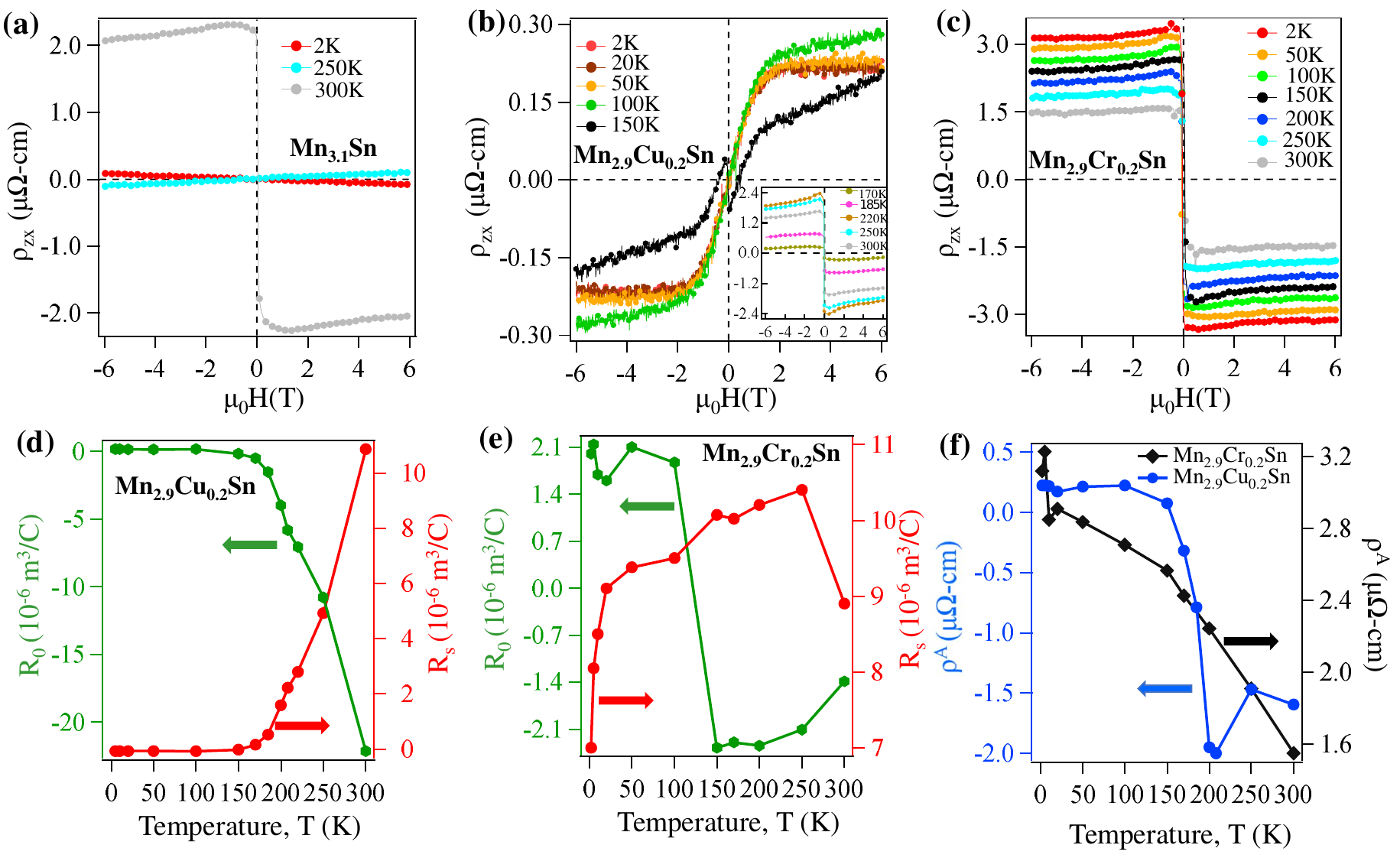}
\caption{Field-dependent Hall resistivity [$\rho_{zx}(H)$] measured from 2\,K to 300\,K for (a)~Mn$_{3.1}$Sn, (b)~Mn$_{2.9}$Cu$_{0.2}$Sn, and
(c)~Mn$_{2.9}$Cr$_{0.2}$Sn. The inset in (b) shows the data taken at the temperatures from 170 K to 300 K for Mn$_{2.9}$Cu$_{0.2}$Sn.  Temperature dependence of the ordinary Hall coefficient $R_0$ and the anomalous Hall coefficient $R_s$ for (d) Mn$_{2.9}$Cu$_{0.2}$Sn and (e) Mn$_{2.9}$Cr$_{0.2}$Sn, respectively. (f)~Temperature dependence of maximum anomalous Hall resistivity for both substituted samples.}
\label{fig5}
\end{figure*}

\begin{figure*}[htbp]
\centering
\includegraphics[width=0.9\linewidth]{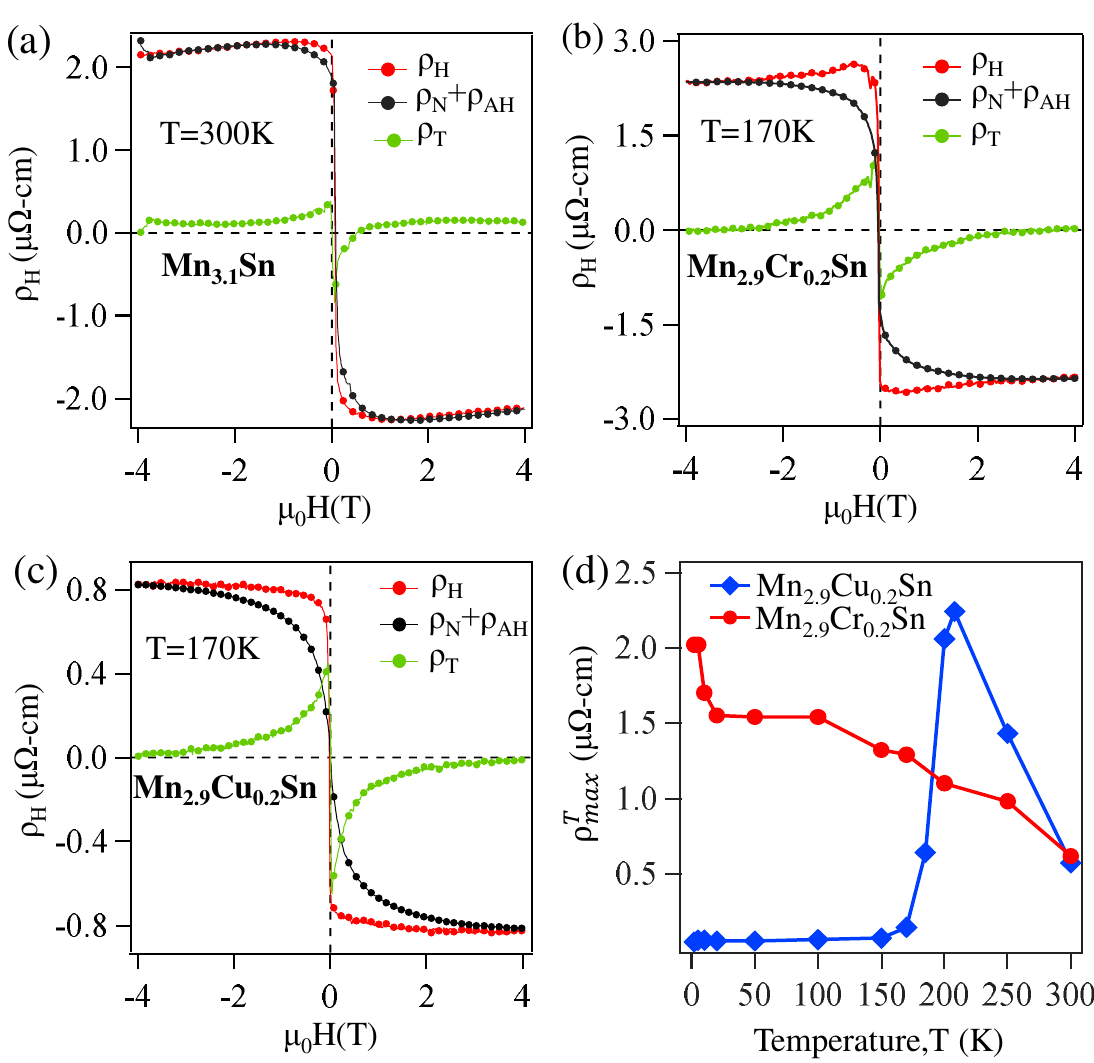}
\caption{(a)- (c)~Field-dependent topological Hall resistivity $\rho_{\mathrm{T}}(H)$ extracted via Eq.~(\ref{eq:THE_extraction}) for Mn$_{3.1}$Sn,
Mn$_{2.9}$Cr$_{0.2}$Sn, and Mn$_{2.9}$Cu$_{0.2}$Sn at representative temperatures.
In each panel, the red curve represent the total experimental Hall resistivity [$\rho_{zx}(H)$],
the black curve represent the fit using the Eq. $\rho_H=\rho_N + \rho_{AH}$ (see the text for more details),
and the green curve represent the extracted topological Hall resistivity ($\rho_{\mathrm{T}}$) using Eq.~\ref{eq:THE_extraction}.
(d)~Temperature dependence of the maximum topological Hall resistivity $\rho_{T}^{max}$ for Mn$_{2.9}$Cu$_{0.2}$Sn and Mn$_{2.9}$Cr$_{0.2}$Sn}
\label{fig6}
\end{figure*}

\begin{figure*}[htbp]
\centering
\includegraphics[width=0.9\linewidth]{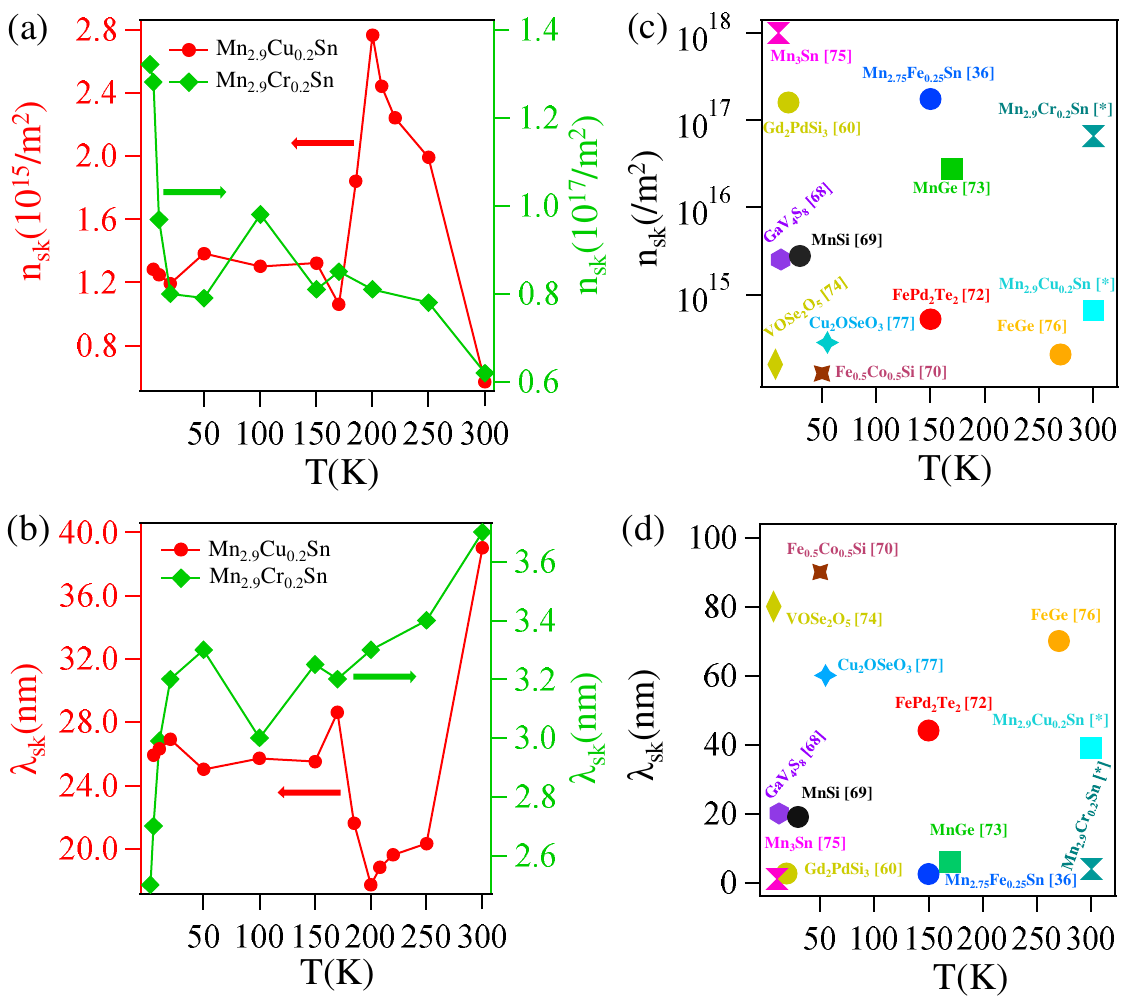}
\caption{Temperature dependence of (a) chiral spin (skyrmion) density ($n_{\mathrm{sk}}$) and (b) the chiral spin (skyrmion) size ($\lambda_{\mathrm{sk}}$) for Mn$_{2.9}$Cu$_{0.2}$Sn and Mn$_{2.9}$Cr$_{0.2}$Sn. Panels (c) and (d)~compile the reported $\lambda_{\mathrm{sk}}$ and $n_{\mathrm{sk}}$, respectively, for various compositions from the literature, together with the corresponding values obtained for the present samples.}
\label{fig7}
\end{figure*}

\subsection{Electrical Transport and Hall Effect Properties}

Fig.~\ref{fig4}(a) presents the temperature dependence of longitudinal electrical resistivity ($\rho_{zz}$) for Mn$_{3.1}$Sn,
Mn$_{2.9}$Cr$_{0.2}$Sn, and Mn$_{2.9}$Cu$_{0.2}$Sn. All three compositions display metallic behavior ($d\rho/dT > 0$) across the
full measured temperature range. The pristine Mn$_{3.1}$Sn exhibits a well-defined $\it{kink}$ near 265 K, associated with the above
discussed spin-reorientation transition (T$_{SR}$)~\cite{sung2018magnetic,low2022tuning, park2025nominal}. Non-magnetic Cu substitution shifts this anomaly to approximately 193 K, reflecting the reduced stability of the in-plane magnetic order upon exchange dilution, whereas
magnetic Cr substitution completely suppresses the T$_{SR}$, producing a monotonically decreasing $\rho_{zz}$ without any anomaly.
The complete suppression of T$_{SR}$ with magnetic element (Fe) substitution was noticed earlier also by the same authors~\cite{low2022tuning}

The temperature dependence of the Hall resistivity ($\rho_{zx}$), measured at $\mu_0H = 1$ T, is shown in Fig.~\ref{fig4}(b).
In Mn$_{3.1}$Sn, $\rho_{zx}$ drops to nearly zero below $T_\mathrm{SR}$, in agreement with earlier reports~\cite{nakatsuji2015large,low2022tuning},
and is understood as the progressive restoration of effective time-reversal symmetry upon entry into the spin-spiral phase,
which quenches the net Berry curvature~\cite{Song2020}. A qualitatively similar but quantitatively distinct behavior is
observed in Mn$_{2.9}$Cu$_{0.2}$Sn. Such as, $\rho_{zx}$ grows with decreasing temperature (negative sign), but below $T_\mathrm{SR}\approx 193\,$K it diminishes, undergoes a sign change from negative to positive below 150 K, and saturates at low temperatures.
This sign reversal is a direct signature of a Berry-curvature redistribution driven by the T$_{SR}$-induced restructuring of the
electronic band topology.

In contrast, Mn$_{2.9}$Cr$_{0.2}$Sn exhibits a monotonic enhancement of $|\rho_{zx}|$ upon cooling, with no sign reversal
or T$_{SR}$-related anomaly. This behavior reflects the persistence of time-reversal symmetry breaking and the associated
Berry curvature across the full temperature range, enabled by the complete suppression of the T$_{SR}$ by Cr substitution.
The noncollinear spin texture in Mn$_{2.9}$Cr$_{0.2}$Sn, which is preserved and reinforced upon cooling, thus maintains
a robust anomalous Hall response throughout. To separately quantify the ordinary and anomalous Hall contributions,
the field-dependent Hall resistivity $\rho_{zx}(H)$ was measured at temperatures from 2 to 300\,K, as shown in Fig.~\ref{fig5}.
The Hall resistivity of a conventional magnetic system is expressed as

\begin{equation}
\rho_{zx}(H) = \mu_0 R_0 H + \mu_0 R_s M,
\label{eq:Hall_conventional}
\end{equation}

where $R_0$ is the ordinary Hall coefficient encoding carrier density and type, $R_s$ is the anomalous Hall coefficient, and $M$ is the magnetization. However, as discussed below, this two-component model is insufficient to account for the full measured $\rho_{zx}(H)$, necessitating an additional topological Hall contribution.

In the case of Mn$_{3.1}$Sn [see Fig.~\ref{fig5}(a)], a clear anomalous Hall signal is observed above
$T_\mathrm{SR}$, in agreement with prior reports~\cite{nakatsuji2015large,Li2023field}. For Mn$_{2.9}$Cr$_{0.2}$Sn
[see Figs.~\ref{fig5}(c) and ~\ref{fig5}(f)], the anomalous Hall resistivity increases steadily upon cooling, reaching a
maximum of $\sim 3.2\,\mu\Omega\,\mathrm{cm}$ at 2\,K. This monotonic enhancement indicates that
Cr substitution preserves and reinforces the Berry-curvature-driven intrinsic anomalous Hall mechanism by stabilizing
out-of-plane spin canting. For Mn$_{2.9}$Cu$_{0.2}$Sn [see Figs.~\ref{fig5}(b) and ~\ref{fig5}(f)], the anomalous Hall response is strongly
temperature-dependent, reaching a maximum of $\sim 2.0\, \mu\Omega\,\mathrm{cm}$ near 208\,K, reversing sign below 150\,K
[see inset of Fig.~\ref{fig5}(b)], and decreasing toward both higher and lower temperatures.
This non-monotonic behavior reflects the strong interplay among spin reorientation, Fermi-surface reconstruction, and
momentum-space Berry curvature redistribution driven by Cu substitution.

The temperature dependencies of $R_0$ and $R_s$, extracted from fits to Eq.~(\ref{eq:Hall_conventional}), are shown in Figs.~\ref{fig5}(d) and~\ref{fig5}(e). In Mn$_{2.9}$Cu$_{0.2}$Sn, $R_0$ remains positive below 150 K, indicating dominant hole-type carrier transport. Above this temperature, $R_0$ changes sign and becomes negative, signaling a crossover to electron-dominated transport. This carrier-type reversal
is a direct transport fingerprint of the temperature-driven electronic structure reconstruction associated with the T$_{SR}$~\cite{chen2021anomalous, Yu2022, Pradhan2023}. Also, see the Supplemental Material to examine the influence of spin configuration on the electronic structure of the parent Mn$_3$Sn and doped-Mn$_3$Sn systems. Simultaneously, $R_s$ increases sharply above 150 K despite a reduction in magnetization, confirming that the anomalous Hall response in this temperature regime is governed by the intrinsic Berry curvature mechanism rather than by the net magnetic moment alone~\cite{nagaosa2010anomalous,onoda2004anomalous}. In Mn$_{2.9}$Cr$_{0.2}$Sn, both $R_0$ and $R_s$ evolve monotonically with temperature, consistent with the absence of a T$_{SR}$ and the preservation of the magnetic ground state~\cite{Liu2020}.

The failure of the conventional two-component model [Eq.~(\ref{eq:Hall_conventional})] to fully reproduce the measured $\rho_{zx}(H)$ in both substituted samples indicates the presence of an additional Hall contribution. The total Hall resistivity is therefore can be written as
\begin{equation}
\rho_{zx}(H) = \mu_0 R_0 H + \mu_0 R_s M + \rho_{T}(H),
\label{eq:Hall_total}
\end{equation}
where $\rho_{T}$ is the Hall resistivity due to the topological Hall effect (THE) arising from the real-space Berry curvature generated by noncoplanar spin textures. The topological Hall resistivity is extracted by subtracting the fitted conventional contributions~\cite{low2022tuning, ghosh2023tuning, low2025topological}:
\begin{equation}
\rho_{\mathrm{T}} = \rho_{zx}(H) - \bigl(\mu_0 R_0 H + \mu_0 R_s M\bigr).
\label{eq:THE_extraction}
\end{equation}

Figs.~\ref{fig6}(a)--\ref{fig6}(c) show the field dependence of the Hall resistivity for Mn$_{3.1}$Sn, Mn$_{2.9}$Cr$_{0.2}$Sn,
and Mn$_{2.9}$Cu$_{0.2}$Sn, respectively. The corresponding topological Hall resistivity, $\rho_{T}$,
extracted using Eq.~\ref{eq:THE_extraction}, is overlaid in each panel. In all three samples, the extracted $\rho_T$ is
 concentrated at low fields and progressively vanishes at high fields, where the magnetic texture is driven toward saturation.
 This field dependence is consistent with a field-suppressible noncoplanar spin texture and its associated real-space Berry curvature.

Further, Fig.~\ref{fig6}(d) shows that, in Mn$_{2.9}$Cr$_{0.2}$Sn, the maximum topological Hall resistivity,
$\rho_{T}^{max}\sim2.02~\mu\Omega\,\mathrm{cm}$, occurs at the lowest measured temperature and
decreases monotonically upon heating. This behavior indicates that the noncoplanar spin texture and the associated real-space Berry
curvature are most strongly stabilized at low temperatures and progressively suppressed by thermal fluctuations.
In contrast, Mn$_{2.9}$Cu$_{0.2}$Sn exhibits a pronounced nonmonotonic temperature dependence,
with $\rho_{\mathrm{T}}^{\mathrm{max}}\sim2.24~\mu\Omega\,\mathrm{cm}$ near 208~K.
The rapid enhancement of the THE between 150 and 208~K coincides with the spin-reorientation regime.
This observation suggests a strong coupling between the temperature-dependent magnetic anisotropy and the emergence
of real-space Berry curvature in the Cu-substituted compound.

The THE in the studied samples arises from a finite scalar spin chirality, $\chi_{ijk}=\mathbf{S}_i\cdot(\mathbf{S}_j\times\mathbf{S}_k)$. This finite scalar spin chirality acts as an emergent magnetic field on itinerant electrons, generating a real-space Berry phase that contributes to $\rho_{\mathrm{T}}$\cite{onoda2004anomalous,kurumaji2019skyrmion, kimbell2022challenges,hao2026enhanced}. Although bulk Mn$_3$Sn is centrosymmetric, chemical substitution locally perturbs the lattice and exchange pathways, thereby modifying the local magnetic interactions and potentially generating finite local Dzyaloshinskii--Moriya interactions (DMI)that favor noncoplanar spin configurations
\cite{grohol2005spin, matan2011dzyaloshinskii,Liu2023, Liu2023a}. In Mn$_{2.9}$Cr$_{0.2}$Sn, the modified magnetic exchange promotes low-temperature spin canting, leading to enhanced scalar spin chirality and a pronounced low-temperature THE. In contrast, Cu substitution modifies the magnetic-anisotropy landscape and stabilizes an evolving noncoplanar spin texture across the spin-reorientation regime, resulting in a broad topological Hall response at intermediate temperatures\cite{Bera2025}.

The contrasting Hall responses of Cr- and Cu-substituted Mn$_3$Sn can be understood in terms of their different effects on the electronic structure and magnetic interactions. In Mn$_{3.1}$Sn, the partially filled Mn $3d$
states support both local-moment formation and itinerant exchange interactions. Substitution of Mn by Cr ($3d^5 4s^1$) introduces a magnetic
dopant with a finite local moment and perturbs the exchange network while largely preserving the connectivity of the Mn kagome lattice
\cite{Liu2020}. Hybridization between the Cr and Mn $3d$ states can enhance the local exchange fields without substantially modifying the carrier density, favoring stronger low-temperature spin canting. Consequently, Mn$_{2.9}$Cr$_{0.2}$Sn retains a predominantly in-plane easy axis, with the in-plane magnetization approximately 1.5--2 times larger than the out-of-plane component, and exhibits a maximum topological Hall resistivity of $\rho_{\mathrm{T}}^{\mathrm{max}}\sim2.02~\mu\Omega$\,cm at low temperature. In contrast, nominally nonmagnetic Cu possesses a filled $3d^{10}$ shell and removes an active magnetic Mn site from the exchange network. The resulting exchange dilution, together with band-filling reconstruction, modifies the itinerant exchange coupling and magnetocrystalline anisotropy, destabilizing the inverse-triangular antiferromagnetic state and driving a spin-reorientation transition at $T_{\mathrm{SR}}\sim193$~K accompanied by an easy-axis reversal \cite{li2023discovery}. The sign change of $R_0$ near $T_{\mathrm{SR}}$ further indicates a substantial reconstruction of the electronic structure. Since the anomalous Hall response of Mn$_3$Sn is highly sensitive to Berry curvature near avoided band crossings and Weyl-related features \cite{Li2023field}, such electronic reconstruction can strongly modify the Hall response. Accordingly, Mn$_{2.9}$Cu$_{0.2}$Sn exhibits a broad, nonmonotonic THE, reaching $\rho_{\mathrm{T}}^{\mathrm{max}}\sim2.24~\mu\Omega$\,cm near 208~K,
rather than being confined to the low-temperature regime.

A comparison with Fe-substituted Mn$_3$Sn further demonstrates that magnetic
dopants can produce markedly different magnetic anisotropy and spin-reorientation behavior. For Mn$_{2.8}$Fe$_{0.2}$Sn
\cite{low2025topological}, a spin-reorientation transition occurs at $T_{\mathrm{SR}}\sim125$~K, while for Mn$_{2.75}$Fe$_{0.25}$Sn it shifts to $\sim115$~K \cite{low2022tuning}. The latter additionally exhibits a weak ferromagnetic-like transition near 240~K and a pronounced enhancement of the out-of-plane magnetic component below $\sim40$~K, resulting in strong low-temperature out-of-plane magnetic anisotropy. The corresponding topological Hall resistivity reaches $\rho_{\mathrm{T}}^{\mathrm{max}}\sim5~\mu\Omega$\,cm at 2~K. With further Fe substitution to Mn$_{2.65}$Fe$_{0.35}$Sn, the spin-reorientation transition is completely suppressed, while a strong ferromagnetic-like transition emerges near 150~K and the low-temperature out-of-plane magnetic anisotropy is further enhanced~\cite{low2022tuning}.                                         In contrast, Mn$_{2.9}$Cr$_{0.2}$Sn shows no distinct $T_{\mathrm{SR}}$ and retains predominantly easy-plane anisotropy, despite the development of low-temperature out-of-plane spin canting. Its maximum THE, $\rho_{\mathrm{T}}^{\mathrm{max}}\sim2.02~\mu\Omega$\,cm, also occurs at low temperature. Thus, although Fe and Cr are both magnetic $3d$ dopants, Fe progressively drives the magnetic ground state toward stronger out-of-plane anisotropy, whereas Cr primarily modifies the exchange interactions and enhances spin canting while retaining the predominantly in-plane easy axis.

The behavior of nonmagnetic Cu substitution is qualitatively different from that of both magnetic dopants. In Mn$_{2.9}$Cu$_{0.2}$Sn, $T_{\mathrm{SR}}\sim193$~K is accompanied by an easy-axis reversal from in-plane to predominantly out-of-plane orientation. More importantly, while the THE in the Fe- and Cr-substituted systems is strongest at low temperatures, the Cu-substituted compound exhibits a broad, nonmonotonic THE within the intermediate-temperature spin-reorientation regime, reaching $\rho_{\mathrm{T}}^{\mathrm{max}}\sim2.24~\mu\Omega$\,cm near 208~K. Therefore, Fe, Cr, and Cu substitutions provide three distinct routes for controlling the magnetic and topological response of Mn$_3$Sn: Fe promotes strong low-temperature out-of-plane anisotropy, Cr preserves the easy-plane anisotropy while enhancing low-temperature spin canting, and Cu induces exchange dilution, easy-axis reversal, and an intermediate-temperature spin-reorientation regime with enhanced THE. This comparison demonstrates that the magnitude and temperature range of the THE are not governed solely by the net magnetization or magnetic anisotropy, but rather by the dopant-dependent interplay among electronic configuration, exchange interactions, magnetic anisotropy, spin reorientation, and the resulting noncoplanar spin configurations.

Next, the magnitude and temperature evolution of the topological Hall effect are intimately governed by the size and density of the underlying noncoplanar spin textures. Chiral or skyrmionic spin structures arise from the competition among the symmetric exchange interaction ($J$), the antisymmetric Dzyaloshinskii--Moriya interaction ($D$), and magnetocrystalline anisotropy ($K$)~\cite{bak1980theory, nakanishi1980origin,kezsmarki2015neel}. In systems hosting a finite skyrmion (chiral spin) density $n_{\mathrm{sk}}$, the Berry phase accumulated by itinerant charge carriers traversing the noncoplanar texture gives rise to an emergent electromagnetic field $B_{\mathrm{eff}}$, producing a topological Hall resistivity~\cite{rout2019field,neubauer2009topological,kurumaji2019skyrmion,han2017scaling} as
\begin{equation}
    \rho_{\mathrm{T}} = P R_0 \, n_{\mathrm{sk}} \, \phi_0,
    \label{eq:THE}
\end{equation}
where $\phi_0 = h/e$ is the flux quantum and
$P = \mu_{\mathrm{sp}}/\mu_{\mathrm{sat}}$ is the conduction-electron spin polarization~\cite{ye1999berry,low2022tuning,liu2026signature}, $\mu_{sp}$ is spontaneous magnetic moment, and $\mu_{sat}$ is saturation magnetic moment.

Since the emergent field $B_{\mathrm{eff}}=n_{\mathrm{sk}}\phi_0$ scales with the effective skyrmion (chiral spin) density $n_{\mathrm{sk}}$, a higher skyrmion (chiral spin) density gives rise to a correspondingly larger topological Hall response~\cite{Kanazawa2011large,kurumaji2017neel}. In the parent compound Mn$_3$Sn, N\'eel-type skyrmions have been associated with a large emergent field of $B_{\mathrm{eff}}\sim3500$~T~\cite{xu2021imaging}, corresponding to a characteristic length scale of approximately 1~nm~\cite{rout2019field}. Using Eq.~(\ref{eq:THE}), we obtain an effective characteristic texture length of $\sim3.7$~nm for Mn$_{2.9}$Cr$_{0.2}$Sn, which is comparable to the characteristic length scale of $\sim2.4$~nm reported for Mn$_{2.75}$Fe$_{0.25}$Sn~\cite{low2022tuning}. In contrast, Mn$_{2.9}$Cu$_{0.2}$Sn yields a substantially larger effective length scale of $\sim39$~nm, comparable to those reported for B20 skyrmion hosts such as MnSi ($\sim18$~nm)~\cite{neubauer2009topological}, FeGe ($\sim70$~nm)~\cite{lebech1989magnetic}, and Cu$_2$OSeO$_3$ ($\sim62$~nm)~\cite{zhang2017direct}, as well as FePd$_2$Te$_2$ ($\sim44$~nm)~\cite{liu2026signature}. These estimates indicate substantially different characteristic length scales for the field-induced noncoplanar spin textures in the Cr- and Cu-substituted compounds. In particular, the much longer effective length scale inferred for Mn$_{2.9}$Cu$_{0.2}$Sn may be indicative of a more extended chiral spin texture in the spin-reorientation regime.

The temperature dependence of $n_{\mathrm{sk}}$ and $\lambda_{\mathrm{sk}}$ for both samples as shown in Figs.~\ref{fig7}(a) and ~\ref{fig7}(b), respectively, reveals a clear dichotomy. In Mn$_{2.9}$Cu$_{0.2}$Sn, $n_{\mathrm{sk}}$ is non-monotonic, reaching a maximum of $2.76\times10^{15}\,\mathrm{m}^{-2}$ around 200\,K. The associated emergent field ($\sim$11\,T) is responsible for the enhanced scalar spin chirality and large THE in the intermediate temperature regime. In Mn$_{2.9}$Cr$_{0.2}$Sn, $n_{\mathrm{sk}}$ decreases monotonically from $1.32\times10^{17}\,\mathrm{m}^{-2}$ at 2\,K, corresponding to $B_{\mathrm{eff}}\sim 550\,$T, and is progressively suppressed by thermal fluctuations similar to that of Mn$_{2.75}$Fe$_{0.25}$Sn~\cite{low2022tuning}.

\section{Conclusions}

In summary, we have demonstrated that magnetic and non-magnetic chemical substitution provide two distinct pathways for controlling noncoplanar spin textures and topological transport in the kagome antiferromagnet Mn$_3$Sn. Cr substitution preserves the predominantly in-plane magnetic anisotropy while modifying the local exchange interactions, leading to enhanced low-temperature spin canting and a pronounced topological Hall response. The corresponding effective texture length scale of $\sim3.7$~nm indicates a relatively short-period noncoplanar magnetic texture. In contrast, non-magnetic Cu substitution strongly perturbs the Mn--Mn exchange network and reconstructs the electronic structure, resulting in a spin-reorientation transition at $T_{\mathrm{SR}}\approx193$~K accompanied by a reversal of the magnetic easy axis. The Cu-substituted compound exhibits a broad topological Hall response that reaches its maximum near 200~K, together with a carrier-type crossover and a substantially larger effective texture length scale of $\sim39$~nm.

The contrasting temperature dependence of the topological Hall response and the inferred texture parameters highlights the distinct roles of exchange interactions, magnetic anisotropy, and electronic structure in stabilizing noncoplanar spin configurations. In particular, the large and non-monotonic topological Hall response in the Cu-substituted compound within the spin-reorientation regime points to enhanced chiral spin correlations and fluctuating noncoplanar textures, rather than providing direct evidence for a static skyrmion lattice. The DFT results further show that Cr and Cu substitution produce markedly different spin-polarized electronic structures and magnetic-anisotropy landscapes, providing a microscopic basis for their contrasting magnetic and transport properties. Overall, our results establish chemical substitution as an effective means of tuning real-space Berry curvature in kagome antiferromagnets and demonstrate that exchange modification, anisotropy control, and electronic-structure reconstruction can be combined to engineer topological Hall responses over widely different characteristic length scales.

\section{Acknowledgement}
S.T. thanks the Science and Engineering Research Board (SERB) and Anusandhan National Research Foundation (ANRF), India, for the financial support through Grant nos. CRG/2023/00748 and ANRF/ARG/2025/007085/PS, respectively.   This research has used the Technical Research Centre (TRC) Instrument facilities of S. N. Bose National Centre for Basic Sciences, established under the TRC project of the Department of Science and Technology (DST), Govt. of India.

\bibliographystyle{elsarticle-num}
\bibliography{Cu_Mn3Sn_Clean}

\section{Supplementary:Doping-dependant Electronic Band Structure}

To further examine the influence of spin reorientation on the electronic structure, we performed first-principles calculations for pristine Mn$_3$Sn and Cr-, Cu-, and Fe-doped Mn$_3$Sn using two representative spin structures  characterized by $M_z=0$ and $M_z=0.5$. The $M_z=0$ configuration represents a compensated noncollinear antiferromagnetic (AFM) state without net in-plane/out-of-plane magnetization, whereas the $M_z=0.5$ configuration represents a canted/noncoplanar state with a finite out-of-plane magnetic component. As shown in Fig.~S1, the doped systems exhibit noticeable changes in the low-energy electronic structure between the $M_z=0$ and $M_z=0.5$ configurations, in addition to the changes induced directly by Cr, Cu, and Fe substitution. In particular, the electronic states near the Fermi level are sensitive to the spin structure, indicating that spin reorientation can induce a reconstruction of the low-energy band structure. Such a reconstruction can modify the relative contributions of electron- and hole-like carriers and may therefore contribute to the experimentally observed variation of the ordinary Hall coefficient $R_0$ across $T_{\mathrm{SR}}$. In contrast, pristine Mn$_3$Sn shows comparatively small changes in its low-energy band structure between the two magnetic configurations. These results suggest that chemical substitution enhances the sensitivity of the low-energy electronic structure to changes in the magnetic configuration.

We emphasize that the present calculations are intended to establish the qualitative influence of spin-reorientation on the electronic structure rather than to provide a quantitative description of the temperature-dependent Hall coefficient. A quantitative first-principles treatment of the doped systems is computationally demanding, as modeling the dilute substitution together with different noncollinear magnetic configurations requires sufficiently large supercells. Moreover, a definitive microscopic description of the abrupt change in $R_0$ across $T_{\mathrm{SR}}$ would require a comprehensive treatment incorporating the temperature-dependent magnetic structure, the associated electronic reconstruction, and carrier transport. The present calculations therefore provide a plausible microscopic basis for the experimentally observed variation of $R_0$, rather than a quantitative reproduction of $R_0(T)$.

\begin{figure*}[ht]
\centering
\includegraphics[width=\linewidth]{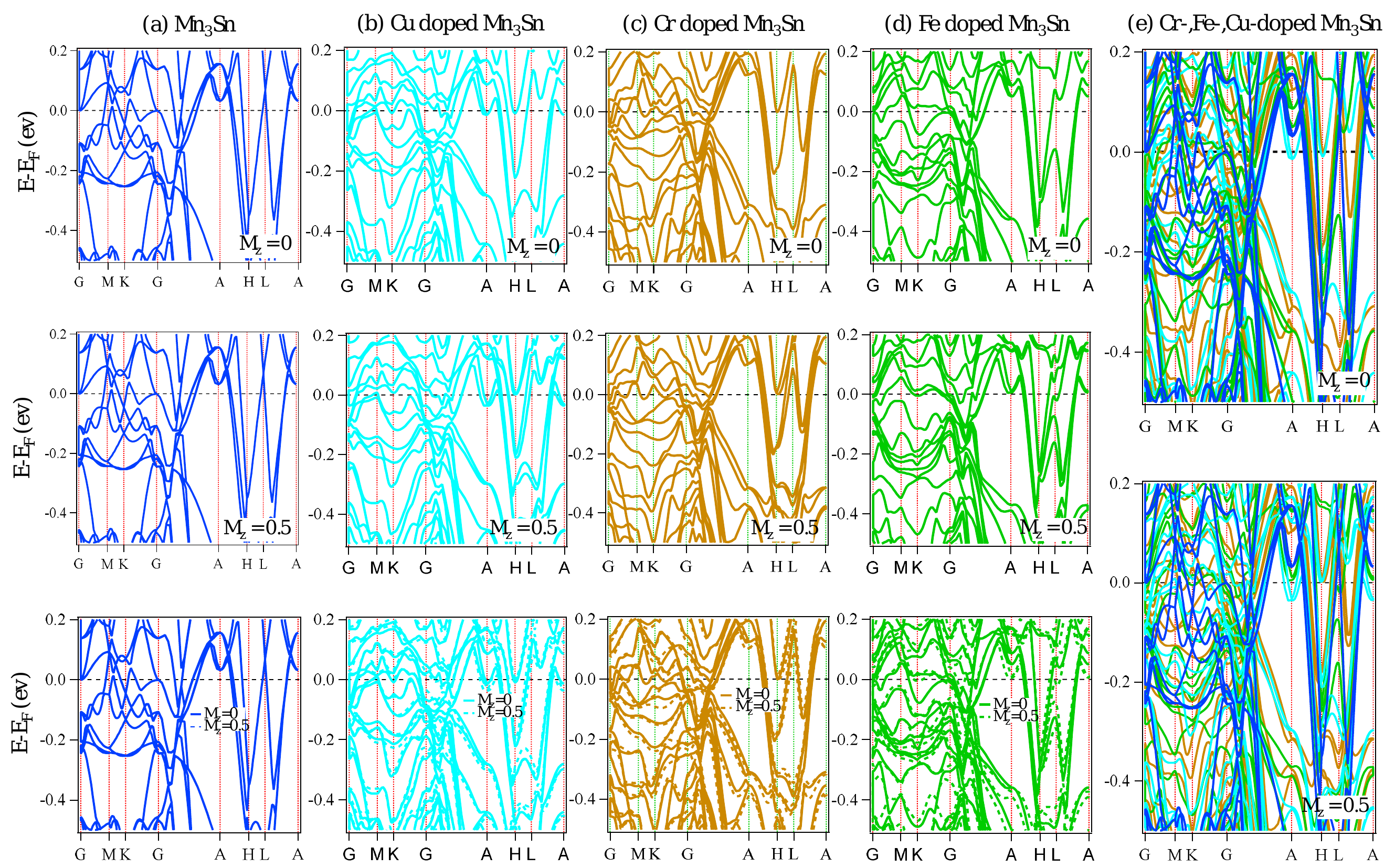}
\caption{Calculated electronic band structures of (a) parent Mn$_3$Sn and (b) Cu-, (c) Cr-, and (d) Fe-doped Mn$_3$Sn for the $M_z=0$ (top panels) and $M_z=0.5$ (middle panels) spin configurations. The bottom panels in (a)--(d) show the corresponding band structures for $M_z=0$ and $M_z=0.5$ overlaid for direct comparison. (e) Overlaid band structures of parent, Cu-, Cr-, and Fe-doped Mn$_3$Sn for $M_z=0$ (top) and $M_z=0.5$ (bottom), illustrating the doping- and spin-configuration-induced modifications of the electronic states near the Fermi level. Details of the computational methods are provided in the main text.}
\label{fig.S1}
\end{figure*}

\end{document}